\documentclass[preprint]{aastex}

\usepackage{graphicx}
\usepackage{psfig}

\slugcomment{Submitted to AJ}

\shorttitle{A gravitationally-lensed Fe LoBAL quasar}
\shortauthors{Lacy et al.}

\begin{document}


\title{The Reddest Quasars II. A gravitationally-lensed FeLoBAL quasar}


\author{Mark Lacy\altaffilmark{1,2,3}, 
Michael Gregg\altaffilmark{2,1}, 
Robert H.\ Becker\altaffilmark{2,1}, Richard L.\ White\altaffilmark{4}, 
Eilat Glikman\altaffilmark{5},
David Helfand\altaffilmark{5} and
Joshua N.\ Winn \altaffilmark{6}}
\altaffiltext{1}{IGPP, L-413, Lawrence Livermore National Laboratory, 
Livermore, CA~94550; gregg@igpp.ucllnl.org, bob@igpp.ucllnl.org}
\altaffiltext{2}{Department of Physics, University 
of California, 1 Shields Avenue, Davis, CA 95616}
\altaffiltext{3}{SIRTF Science Center, Caltech, Mail Code 220-6,
Pasadena, CA 91125; mlacy@ipac.caltech.edu}
\altaffiltext{4}{Space Telescope Science Institute, Baltimore, MD~21218;
rlw@stsci.edu}
\altaffiltext{5}{Columbia University, Department of Astronomy, 550 West 120th 
Street, New York, NY 10027; eilatg@astro.columbia.edu, djh@astro.columbia.edu}
\altaffiltext{6}{Harvard-Smithsonian Center for Astrophysics, 60 Garden
Street, Cambridge, MA 02138; jwinn@cfa.harvard.edu}
\begin{abstract}
We report the discovery of a $z=2.65$ low-ionization iron broad
absorption line quasar, FIRST J100424.9+122922, which is
gravitationally-lensed by a galaxy at $z\approx 0.95$.  The object was
discovered as part of a program to find very red quasars by matching
the FIRST radio survey with the 2-MASS near-infrared survey.
J100424.9+122922 is the second lensed system to be found in this
program, suggesting that many gravitational lenses are probably missed
from conventional optical quasar surveys.  We have made a simple 
lens model and a rough estimate of the reddening in the immediate
environment of the quasar which suggests that the quasar is
intrinsically very luminous and is accreting at close to the Eddington
limit of its $\sim 10^9 M_{\odot}$ black hole.  The lensing galaxy has
a small amount of dust which is responsible for some excess reddening
observed in the fainter image of the quasar, but is otherwise a fairly
typical massive elliptical galaxy.  We model the selection effects working
against the detection of red quasars in both lensed and unlensed
samples.  We show that these selection effects are very effective at
removing even lightly-reddened high redshift quasars from
magnitude-limited samples, whether they are lensed or not.  This
suggests that the red quasar population in general could be very
large, and in particular the class of iron broad absorption line
quasars of which J100424.9+122922 is a member may be much larger than
their rarity in magnitude-limited samples would suggest.  
\end{abstract}



\section{Introduction}

We are currently conducting a survey for very red objects selected
using the Two-micron All Sky Survey (2MASS; Kleinmann et al.\ 1994)
and the Faint Images of the Radio Sky at Twenty-centimeters survey
(FIRST; Becker, White \& Helfand 1995; White et al.\ 1997 ). We select 
stellar infrared objects coincident with FIRST radio sources which lack
(or have only faint) counterparts on the Palomar
Observatory Sky Survey plates.  To date we have examined about 100 
candidate red quasars, most using Keck or Lick spectroscopy. Of these, 
$\approx 13$ have optical spectra characteristic of $z>0.5$ quasars reddened 
by dust (final classification will need to await near-infrared spectroscopy
as broad lines sometimes appear in the near-infrared where the extinction to 
the quasar broad line region is lower). 
In a previous paper (Gregg et al.\
2002) we reported the discovery of two dust-reddened quasars
discovered as part of this survey -- FIRST J013435.6-093102 and FIRST
J073820.1+275045 -- one of which (J0134-0931) is gravitationally lensed
(see also Winn et al.\ [2002]). In this paper we describe the recent
discovery of a second gravitationally-lensed system,
FIRST~J100424.9+122922 (hereafter J1004+1229).  This quasar is
reddened by a combination of dust and extreme, low ionization broad
absorption in the rest-frame ultraviolet, making it a member of the
rare sub-class of low-ionization broad absorption line quasars, dubbed
``FeLoBALs'' by Becker et al.\ (1997), as much of the absorption is due
strong absorption from
metastable \ion{Fe}{2} (Hazard et al.\ 1987; Becker et al.\ 1997).

There are two main hypotheses concerning the nature of 
broad absorption line quasars (BALs). One is that most quasars possess
BAL flows, but we only see them if the quasars are 
viewed at a particular angle to the line of sight which grazes a torus
of gas around the nucleus (hereafter the ``orientation 
hypothesis'', e.g.\ Weymann et al.\ 1991). The other, which applies
particularly for the case of the low-ionisation BALs [LoBALs], is 
that BALs are quasars
at an early stage in their evolution (hereafter the ``youth hypothesis'', 
e.g.\ Voit, Weymann \& Korista 1993).  Recently, support for the youth 
hypothesis has come both from
an analysis of the radio properties of BALs in FIRST (Becker et al.\
2000), which indicates a range of orientations can produce BAL quasars, 
and also an apparent
association of BALs with infrared-luminous quasars having hosts with merging
morphologies and host spectra with ``post-starburst'' features
(Canalizo \& Stockton 2001).  A further, but more controversial, hint 
that BALs may represent a stage in the life cycle of a quasar is that BALs 
may have steep intrinsic
X-ray spectral slopes (Mathur et al.\ 2001; but see 
also Gallagher et al.\ 2001 who argue that the X-ray spectral slopes of 
BALs are not significantly steeper than those of normal quasars).
Steep X-ray slopes may be connected with high accretion rates relative
to the Eddington limit (Laor et al.\ 1997), so if BALs do indeed have 
steeper X-ray slopes this would suggest a link between 
high accretion rates and the BAL phenomenon.  If it is assumed that high
accretion rates are typical in the early life of a quasar, perhaps
because the supply of fuel capable of falling into the black hole is
greatest just after the triggering event, then a consistent picture
emerges in which BAL flows are a symptom of high accretion rates
produced as radiation pressure from a young quasar close to the
Eddington limit pushes away excess material.

The first FeLoBAL to be identified was LBQS~0059-2735 (Hazard et al.\
1987). Since then five more have been found in the FIRST Bright Quasar
Survey (FBQS) (Becker et al.\ 1997; 2000).  These objects are
overlooked in quasar surveys based on UV/optical color or objective
prism selection, not only because of their extreme absorption features
and lack of prominent emission lines, but also because, like other
LoBALs, they tend to be more heavily reddened by dust
than the normal quasar population (Sprayberry \& Foltz 1992).  Like
normal BALs, few are very radio-loud, so they do not appear in surveys
with high radio flux density limits either.  Quasar surveys based on
the FIRST radio survey, are, however, better suited to finding such
objects in significant numbers, because of the faint radio flux limit
and the lack of the necessity for candidate selection based on
exceptionally blue colors.

For our other FIRST/2MASS gravitational lens, J0134-0931, we were
unable to determine whether the reddening arose predominately in the
immediate environment of the quasar, or in the lensing
galaxy. Determining the source of the reddening (and extinction) is
critical, however: if heavy reddening by lenses is commonplace, then
many gravitationally lensed systems may be missing from
optically-based lens surveys.  If, on the other hand, 
the reddening is in the host, then
many red quasars may be missing from current quasar surveys, with
important implications for attempts to match the mass accreted onto
black holes during the ``quasar epoch'' to the masses of black holes
in galaxies today (e.g.\ Merritt \& Ferrarese 2001).  An association
between BAL quasars and gravitational lensing was noted by Chartas
(2000), suggesting that reddening in the hosts usually dominates, but
examples of lensed quasars reddened by their lenses are also known
[B0218+357 (e.g.\ Menten \& Reid 1996), and PKS 1830-211 (e.g.\
Wiklind \& Combes 1996)].

Our infrared imaging and optical and infrared spectroscopy demonstrate
conclusively that J1004+1229 is a lensed FeLoBAL and allow us to
derive the basic quantities of the lensing system and also investigate
the nature of the lensing galaxy and quasar, as detailed in Sections
2-4.  J1004+1229 is the second
gravitationally lensed, $z>0.5$ red quasar found in our survey.  
This lensing rate
(2/13) is much larger than higher that for the overall quasar population, 
though the statistics are poor so far.  In Section 5 we 
model the effects of lensing and reddening on the optical and
near-infrared selection of quasars, attempting to account for this
apparently high lensing rate of reddened quasars.

We assume a cosmology with $\Omega_{\rm M}=0.3, \Omega_{\Lambda}=0.7, H_0=60
{\rm kms^{-1}Mpc^{-1}}$, unless otherwise stated.  

\section{Observations}

\begin{figure}
\plotone{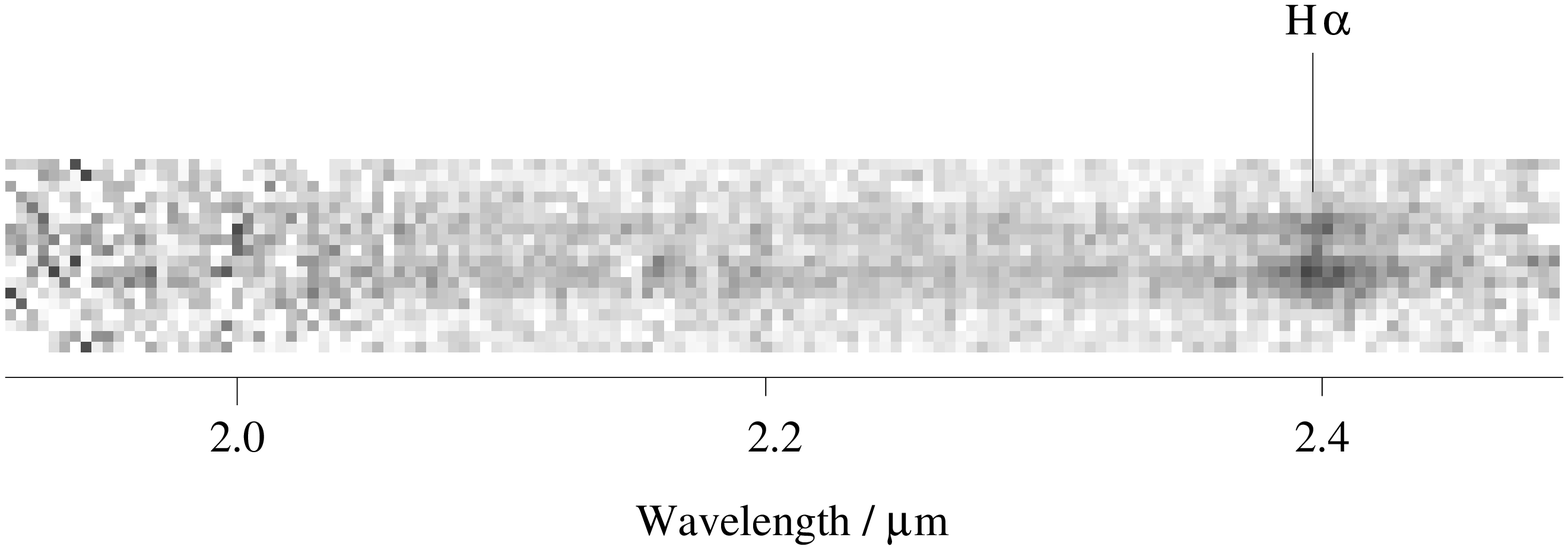}
\caption{$K$-band spectrum of J1004+1229, obtained with NSFCAM at the
IRTF, showing the H$\alpha$ emission line in the two components A (lower) and
B (upper).}
\end{figure}

\begin{figure}
\plotone{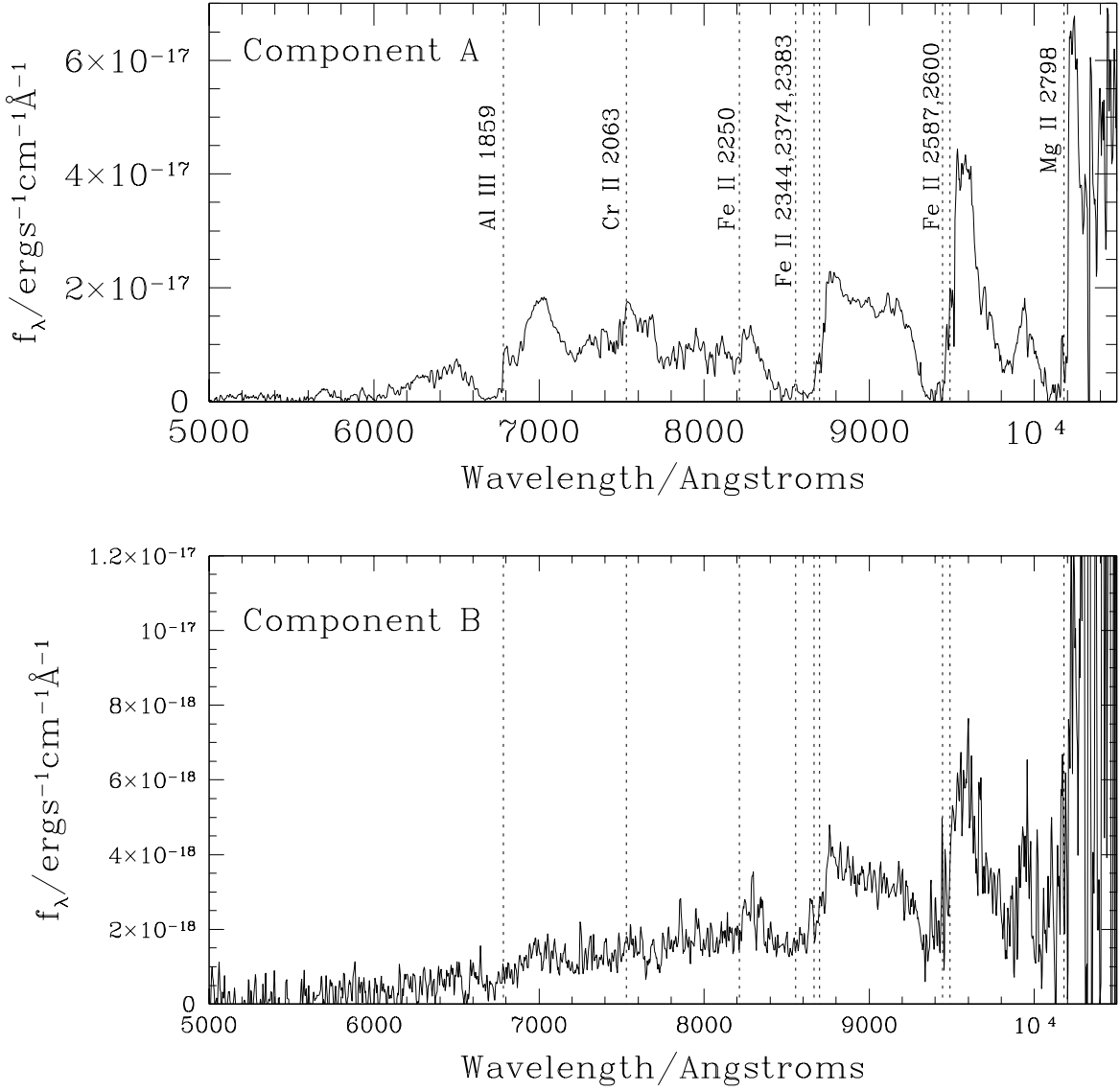}
\caption{The spectrum of components A and B of J1004+1229. The spectral 
resolution in these smoothed spectra is 8\AA. The dotted lines, labelled in 
the upper panel,
mark the redshifted wavelengths of some of the stronger 
broad absorption features in the spectrum.}
\end{figure}

\begin{figure}
\plotone{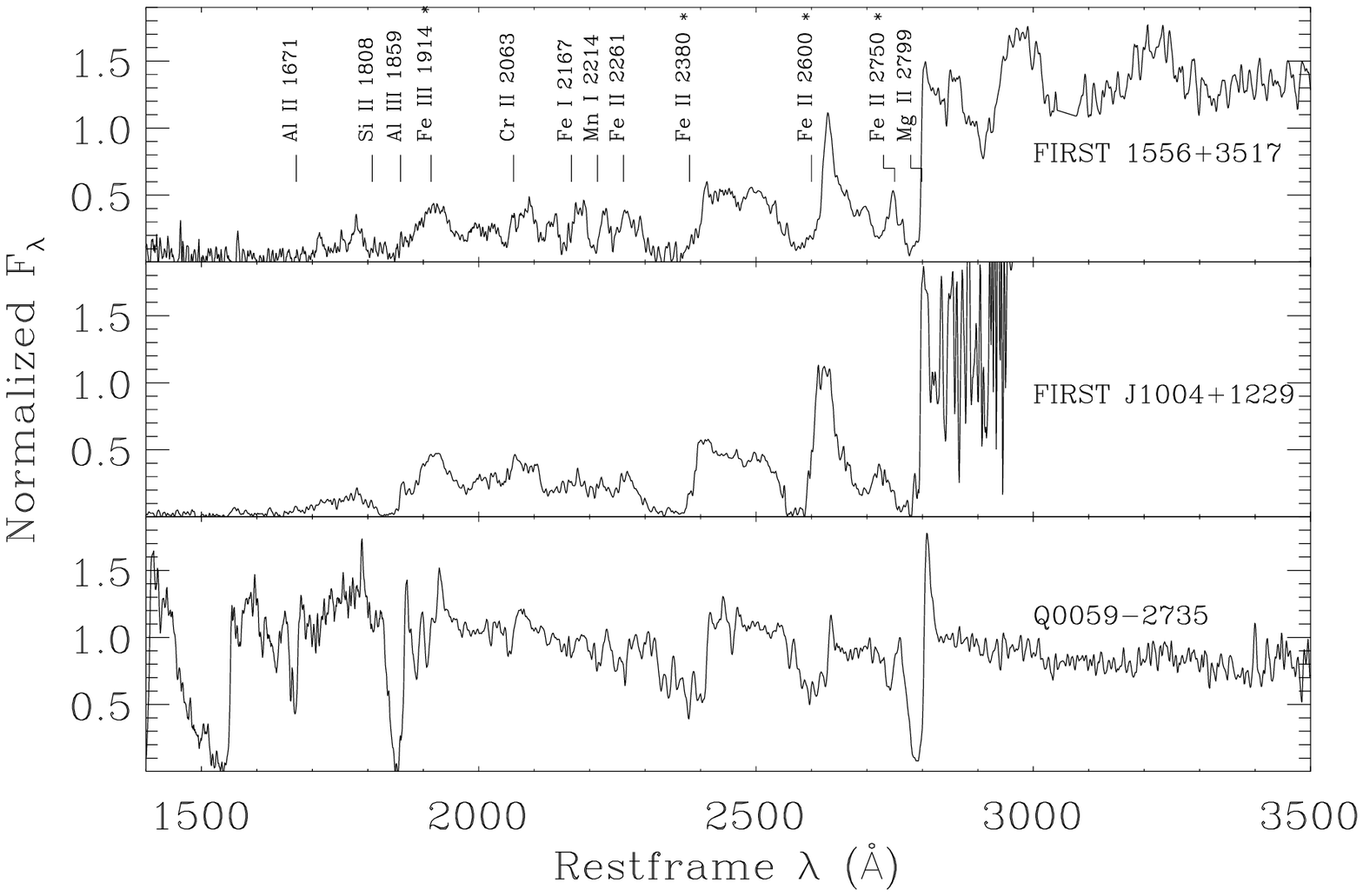}
\caption{The spectrum of J1004+1229A compared to that of the FIRST
FeLoBAL J1556+3517 and the original FeLoBAL, Q0059-2735 (Hazard et
al.\ 1987).  The similarity of J1004+1229 to J1556+3517 is
particularly strong.  Prominent absorption features typically
found in FeLoBALs are marked; the
asterisk denotes a metastable state.}
\end{figure}

Matching of FIRST and 2MASS showed that the 
$K^{'}=14.5$ 2MASS source corresponding to J1004+1229
is 0.3 arcsec from the 12mJy FIRST radio source (compared to estimated
rms position errors of 0.2 arscec in 2-MASS and 0.5 arcsec in 
FIRST), but that this 
object was not present on second-generation digitised sky survey plates to 
a limit of $R\approx 20.8$.
In this Section, we describe observations of J1004+1229 in the radio, 
near-infrared and optical bands. Broad band magnitudes and radio flux 
densities for the system are listed in Table 1.

\subsection{Spectroscopy}
 
J1004+1229 was initially identified as an FeLoBAL in a Keck ESI
spectrum taken on 2000 December 29, though the gravitationally lensed
counterimage was not noticed at the time.  

Infrared spectroscopy with
the NASA Infrared Telescope Facility (IRTF) using NSFCAM was taken on
2001 April 16 (UT) using the $K$-band grism.  The spectrum was nodded
up and down the slit in the standard two-point ``ABBA'' pattern (Eales
\& Rawlings 1993) to facilitate sky subtraction.  A $K$-band
acquisition image showed two components separated by $\approx
1.5$-arcsec which were fortuitously almost aligned with the
north-south slit.   The spectra show that both components have an
emission line at an identical redshift (Figure 1).  We denote the
bright image as component A, and the faint image as component B
throughout this paper.

A further ESI spectrum was taken on 2001 May 26 with the 1-arcsec
slit rotated to a sky PA of 9 deg.\
to include both images of the quasar (Figure 2).  Two 900s
integrations were taken, and both components of the lensed system were
easily resolved along the slit in the $0\farcs5$ arcsec seeing. The 
data were calibrated using an observation of Fiege 34.

Figure~3 compares the ESI
spectrum of J1004+1229 with two other FeLoBALs, LBQS~0059-2735 (Hazard et al.\
1987) and
FIRST~1556+3517 (Becker et al.\ 1997).  The similarity of J1004+1229
and FIRST~1556+3517 is evident. Further discussion of the spectrum 
may be found in Section 3.1. The spectral resolution of 
the observations is $\approx 2$\AA, though the spectra have been smoothed
to 8\AA$\;$resolution in Figures 2 and 3.


\begin{figure}
\plotone{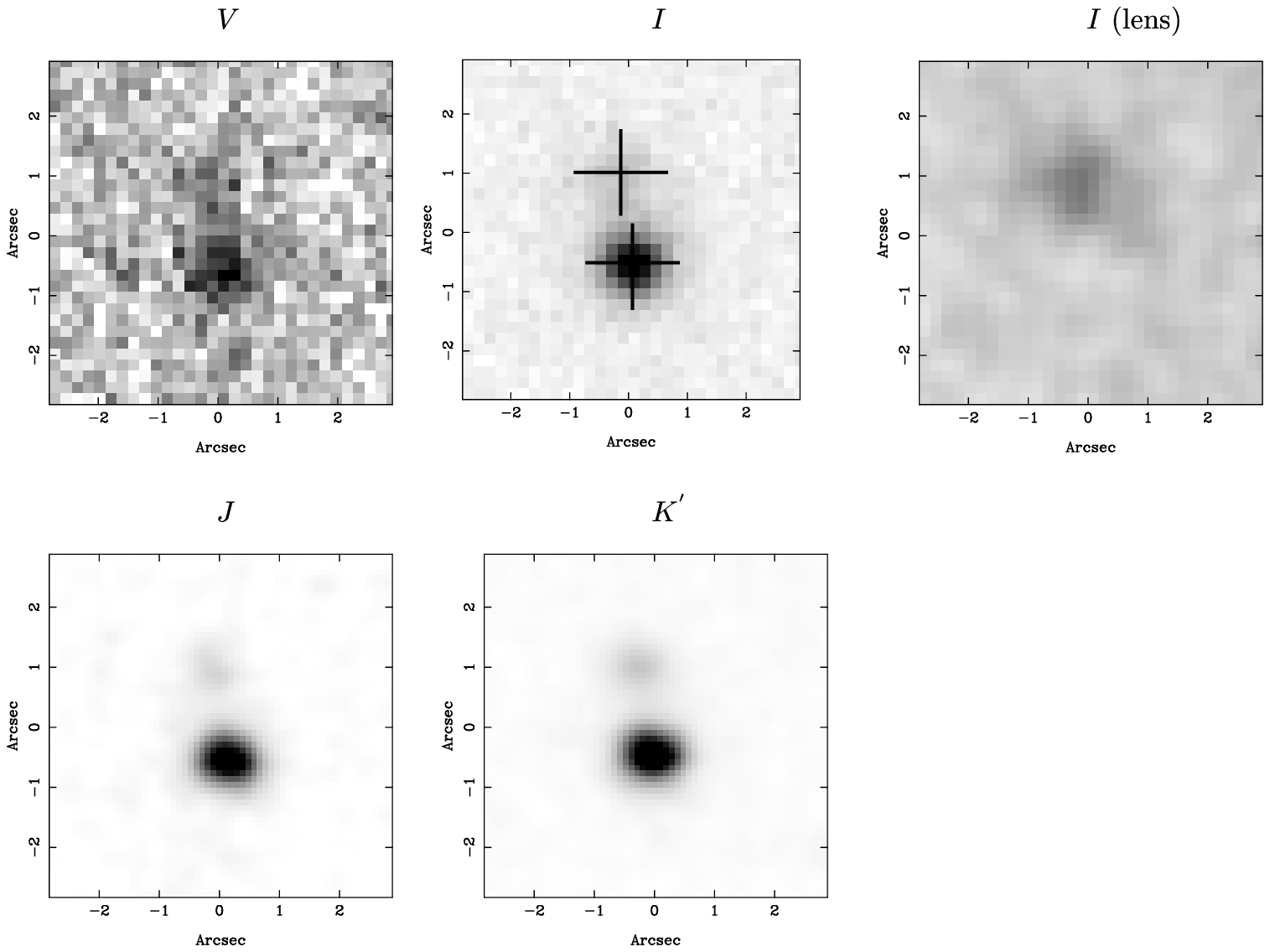}
\caption{Images of J1004+1229. Top row, left to right: $V$, $I$, and the 
$I$-band image with our best estimate of the the two lensed quasar images 
subtracted to reveal the lensing galaxy. Bottom row, left to right:
$J$ and $K^{'}$. Crosses on the $I$-band image mark the positions of the 
radio components detected with the VLBA (see Figure 5).}
\end{figure}


\subsection{Optical and near-infrared imaging}

Infrared images in $J$ and $K^{'}$ were taken with NSFCAM at the IRTF
on 2001 April 16. The data were taken in a 9-point mosaic pattern with
60s exposures at each point, and were reduced using the {\sc dimsum}
package after flat-fielding with dome flats.  The weather was
non-photometric and the seeing 0\farcs7.  
$V-$ and
$I-$band images were obtained with the Andalucia Faint Object
Spectrograph Camera (ALFOSC) on the Nordic Optical Telescope (NOT) on
2001 April 22 for $3\times 300$s and $3 \times 600$s respectively, and
flat-fielded with twilight flats.  Both images show the counterimage
component to be extended, most probably due to blending with the
lensing galaxy.  The images are shown in Figure~4, where we have also
attempted to subtract our best estimate of the contribution of the
quasar images from the $I$-band image (see Section 3.2) to reveal the
lensing galaxy.  
Table~1 lists the broad-band magnitudes, flux densities and
image-to-counterimage ratios.

\subsection{Radio imaging}

We observed J1004+1229 at 1.67 GHz with the VLBA\footnote{ The Very
Long Baseline Array (VLBA) is operated by the National Radio Astronomy
Observatory, a facility of the National Science Foundation operated
under cooperative agreement by Associated Universities, Inc.} on 8--9
October 2001, using all 10 antennas.  We used the compact
radio source J1002+1216 as the phase reference (26\arcmin away,
peak flux density 0.13~Jy), with a switching time of 5 minutes.  The
total dwell time on J1004+1229 was 200 minutes.
 
The data were recorded in eight 8~MHz bands in dual-polarization mode,
for a total observing bandwidth of 32~MHz per polarization.  Each
8~MHz band was subdivided into $16\times 0.5$~MHz channels.  The
sampling time was 1~s.  For the correlation phase centers, we used the
JVAS position (J2000) of the phase reference source ($10\fh 02\fm 52\fs 8457$,
$+12\fdg 16\fm 14\farcs588$; Browne et al.\ 1998), and the
FIRST position of the target source. 

Calibration was performed with standard AIPS tasks.  
We used a fringe-fitting interval of 2 minutes.
Prior to imaging, we averaged
the data in time into 6~s bins and in frequency into 1~MHz bins.  This
level of sampling was sufficient to reduce bandwidth and time-average
smearing to less than 1\% within $1\farcs5$ of the phase center.
Imaging was also performed with AIPS.  We created a 
$2\arcsec\times 2\arcsec$ image centered on the
expected A--B midpoint.  After component A was subtracted using the 
CLEAN algorithm, a
significant peak ($\sim 9\sigma$) was apparent in the residual image.
This peak is the brightest peak in the residual image, it does not
belong to the sidelobe pattern of component A, and it is near the
expected location of component B.  We concluded that it represents
component B of the lensed system.
The final contour maps, shown in Figure~5, were produced after
phase-only self-calibration with a 60 minute solution interval.
 
When A and B are modeled as elliptical Gaussian components, their flux
densities are $6.23$~mJy and $0.99$~mJy, respectively.  The noise on the
map is $0.11$~mJy~beam$^{-1}$, and the the flux
density ratio (A/B) is $6.3\pm 0.7$.
The overall flux density calibration is uncertain by 5\%.  
Component A is $263.3\pm 1.0$~mas
west and $1517.2\pm 1.7$~mas south of component B.  The position of A
is (J2000) $10\fh 04\fm 24\fs 8858$, $+12\fdg 29\fm 22\farcs313$.
The lowest surface-brightness contours of component A suggest the
source is partially resolved.  The elongation is in a direction nearly
perpendicular to the A/B separation, which is suggestive of the
tangential stretching that is characteristic of gravitational lensing.
However, it is possible that the elongation is an artifact of residual
gain errors rather than intrinsic source structure.
Further evidence for extended emission comes from the difference between 
the 1.4 GHz flux density from FIRST of 12.3mJy, measured in an 
5.6-arcsec beam, and the 7.2 mJy total flux measured 
at 1.67 GHz in the VLBA image. The source has a flux density at 1.4 GHz
of 11.8 mJy in the $\approx 45$-arcsec resolution NVSS survey 
(Condon et al.\ 1998), identical to the FIRST flux 
density to within the errors.
Thus assuming the difference in flux density between the VLA and VLBA 
measurements 
is not due to variability, this suggests the VLBA has resolved out 
extended emission from the radio source, and also that the source is not 
significantly larger than the FIRST beamsize.

\begin{figure}
\plotone{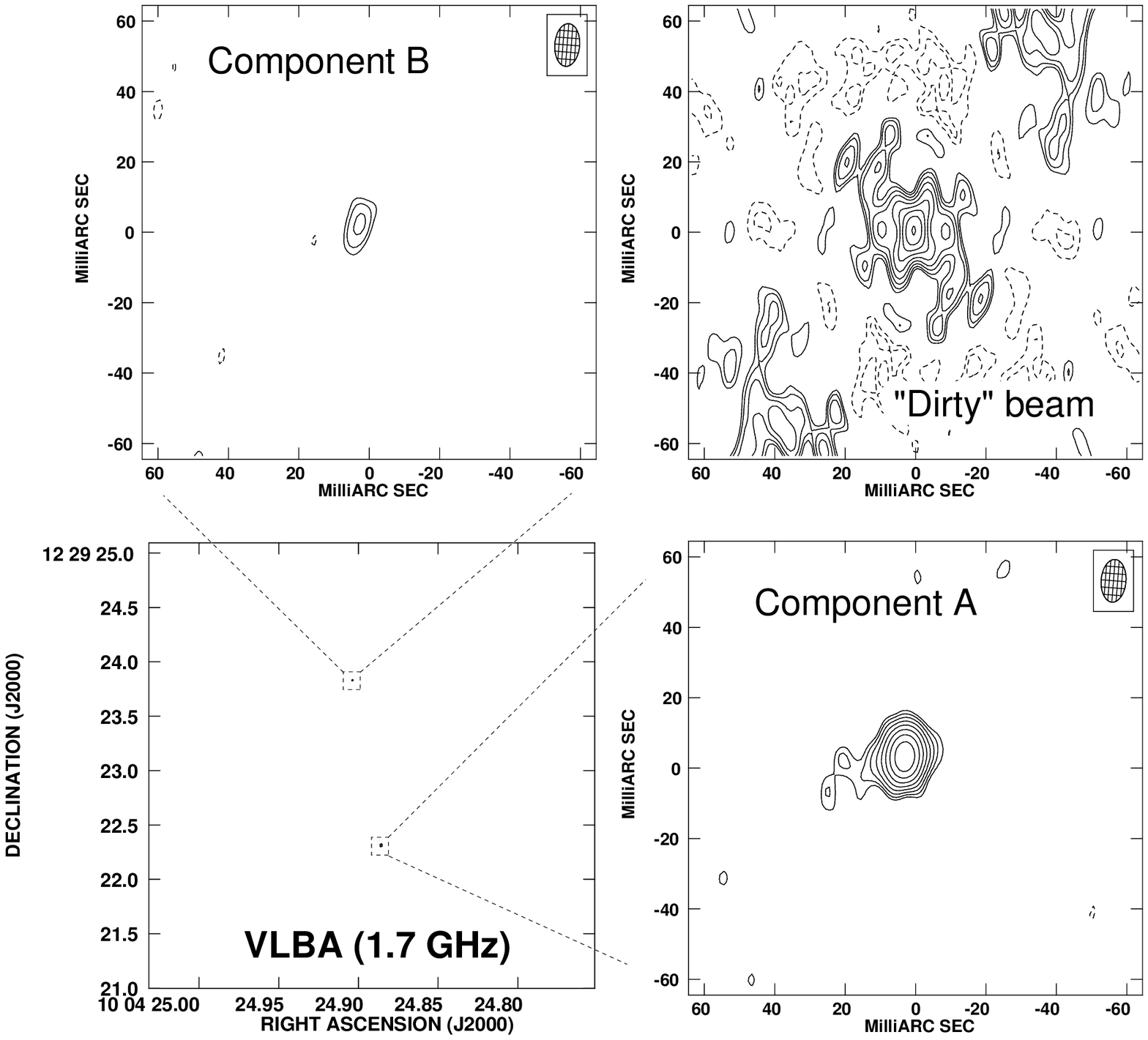}
\caption{The 1.7 GHz VLBA image of J1004+1229. The bottom left panel 
shows the whole map, with the bottom right and top left panels showing
close-ups of components A and B respectively. The dirty beam is shown 
in the top-right panel.}
\end{figure}

\section{Results and Interpretation}

\begin{table}
\begin{tabular}{lcc}
Band & Mag/Flux & A/B ratio\\
$V$  & $22.9\pm 0.1$  & $\sim 4$ \\
$I$  & $19.7\pm 0.1$  & $4.1 \pm 0.5$\\
$J$  & $16.55\pm 0.16$& $7.7\pm 0.2$\\
$K^{'}$  & $14.54\pm 0.09$&$6.8 \pm 0.2$\\
1.4 GHz& 12.3 mJy&- \\
1.67GHz&  -      & $6.3 \pm 0.7$\\
\end{tabular}

\caption{Photometry and image ratios for the J1004+1229
system. Near-IR photometry is from the 2-MASS catalogue, optical
photometry from our NOT observations and the radio flux density from FIRST.
Image-to-counterimage ratios were determined from PSF fitting using
DAOPHOT on our NOT and IRTF data in $I$, $J$ and $K^{'}$.  
The signal-to-noise ratio in the 
$V$-band image was too low to permit a reliable
measurement of the A/B flux ratio.}
\end{table}

\begin{table}
\begin{tabular}{lcc}
Quasar redshift & $2.65 \pm 0.01$\\
Quasar $M_B$ & $-28.5 +2.5{\rm lg}(\mu/3) - 2.5 {\rm lg}(f_B/3)$ \\
1.4GHz radio luminosity & $2 \times 10^{25} \times (3/\mu) {\rm WHz^{-1}sr^{-1}}$\\
Lens redshift & $0.95\pm 0.01$ \\
Lens $M_B$ & $\approx -21.1$ \\
Lens $\sigma_v$ & $\approx 240 {\rm kms^{-1}}$\\
Lens $M/L_{B}$  & $\approx 11 h M_{\odot}/L_{\odot}$ \\ 
\end{tabular}

\caption{Derived quantities for the J1004+1229 system. 
The lensing magnification of the quasar is
denoted $\mu$, and dust attenuation factor in rest-frame $B$-band
$f_B$.  The radio flux is assumed to originate entirely from the
quasar. The mass-to-light ratio is quoted for 
$H_0=100h {\rm kms^{-1}Mpc^{-1}}$}
\end{table}

\subsection{The quasar spectrum}


Our ESI spectra of J1004+1229 show a complicated absorption line
spectrum with an apparent edge at 10200\AA, which we ascribe to Mg{\sc
ii} absorption at $z=2.65$ (Figures 2 and 3).  This interpretation is
confirmed by the $K$-band spectrum, which shows a strong, broad (3500
kms$^{-1}$) emission line at 2.40$\mu$m, corresponding to H$\alpha$ at
$z=2.66\pm 0.01$.  Most of the absorption features can be tentatively
identified with low-ionization species; for example, Mg{\sc ii} 2796/2803 and
Al{\sc iii} 1859 are clearly present.  Much of the absorption between
these wavelengths is due to metastable excited states of Fe~II and 
perhaps Fe~III (Hazard et al.\ 1987; Becker et al.\ 1997), hence its
classification as an FeLoBAL.  The absorbing systems
which give rise to the spectral features seem to be quite complex, making
the resemblance between J1004+1229 and FIRST~1556+3517 particularly
striking (Figure~3). Nevertheless, there are detailed differences in the 
spectra, for example the Fe~I 2167 line, apparently present in FIRST~1556+3517
is absent from J1004+1229. The blueshifts extend to 
$\approx 10000\, {\rm kms^{-1}}$ even for the low ionization lines.  
J1004+1229A has very black absorption troughs,
indicating a high covering factor for the absorbing clouds.

\subsection{Evidence for lensing}

The infrared and optical spectra, which indicate that both 
components A and B are members of the rare FeLoBAL class of quasar, taken 
together with 
the similarity of the image-to-counterimage ratio in the near infrared and 
the radio indicate that components A and B are 
almost certainly lensed images of a single quasar.  The 
optical spectra of A and B do, however, differ in detail, and in this section
we discuss whether these differences are consistent with lensing.

In a simple, isolated three-image lens system, the brightest component 
is farthest from the lensing galaxy, and the faintest (demagnified) 
component is closest. As third images are only rarely detected we modelled the
system as containing two images, and assumed that the likely effects
of lensing on the spectrum were blending of the weaker component with
light from the lensing galaxy, and possibly differential reddening of
the two images by dust in the lens.
This picture is borne out by the photometry (Table 1): the
image-to-counterimage ratio increases from the radio to $K^{'}$ and from 
$K^{'}$ to $J$, as one
would expect if the counterimage is more strongly reddened, but then
decreases again in $I$ as light from the lensing galaxy is blended
into that from the counterimage. 

We thus attempted to decompose the spectrum of component B into a
reddened version of component A plus the lensing galaxy.  Fortunately
the large number of strong features in the spectrum of the quasar
makes this relatively easy to do. The unknown nature and redshift of
the lensing galaxy mean that a large parameter space would need to be
covered in a full analysis; we have therefore decided to restrict
ourselves to a fairly course grid of reddenings to
simplify the problem. 

We explored reddening values of $E(B-V)$ from 0 to 0.5 in steps of
0.1, assuming a ratio of selective to total extinction, $R_V=3.1$ and
a Milky Way extinction law (Cardelli, Clayton \& Mathis 1989) (Figure~6).
Initially we assumed a lens redshift $z_{\rm L}=1$ at which we applied
the extinction.  We fixed the true image:counterimage ratio by
estimating the excess reddening of component B at $K$-band and then
divided the $K$-band image:counterimage ratio by this excess reddening
factor.

An $E(B-V)\approx 0.2$ was found to best remove the quasar features from the
spectrum of B, leaving a residual which is smooth apart from a break
at 7800\AA. We interpret this as a 4000\AA$\;$break at $z=0.95$. 
In the lower panel of Fig.\ 6 we show the residuals from subtracting 
normalized, reddened spectra of component A from that of 
component B for reddenings of $E(B-V)=0.1,0.2$ and $0.3$ at the lens redshift.
The intrinsic flux ratio we obtain for the best-fit, namely 5.5 is 
a little lower than the 6.3 obtained from the VLBA data (though only by 
1.2-$\sigma$). Inspection of Figure~6 
seems to indicate that the intrinsic flux 
ratio lies between $5$ and $6.2$ (corresponding to reddening by the lens 
between $E(B-V)=0.1$ and $0.3$), but spatially-resolved spectroscopy
of the lens is required to make a more accurate estimate.

As the resolution of our images is insufficient to show the lens resolved 
from component B, we adopted a simple singular isothermal sphere model
with which to extract the rough physical parameters of the lens 
system. In this model, the image to counterimage ratio is given 
by:
\[ R=\frac{\alpha + \theta_{\rm s}}{\alpha - \theta_{\rm s}}, \]
where $\theta_{\rm s}$ is the angle between the source position and the
lens center, and $\alpha$ is the deflection angle induced by the lens
(which for this model is independent of impact parameter). The 
bright and faint images
are offset by $\alpha + \theta_{\rm s}$ and $\alpha - \theta_{\rm s}$
from the lens center, respectively. 
Thus given the image:counterimage ratio of 
5.5 and separation ($2\alpha$) of 1.54 arcsec, we expect the fainter image to 
be offset by 0.24 arcsec from the lens, consistent with us failing to
resolve them in our images. The total magnification in this model is 
$2 \alpha / \theta_{\rm s}=2.9$. 

We derive a velocity dispersion for the lens of $\approx 240 {\rm km \,
s^{-1}}$, a mass within the Einstein radius of $\approx 3.0 \times
10^{11} M_{\odot}$, and a blue mass-to-light ratio of $\approx 11 h $
in solar units (where $H_0=100h {\rm kms^{-1}Mpc^{-1}}$), 
all of which are consistent with the lens
being a massive elliptical galaxy. The optical magnitude estimated from the 
spectra and the $I$-band image is also consistent with this, and suggests that
the galaxy has a stellar 
luminosity about 0.5 magnitudes brighter than $M_B^*$ at zero redshift.

\onecolumn

\begin{figure}
\plotone{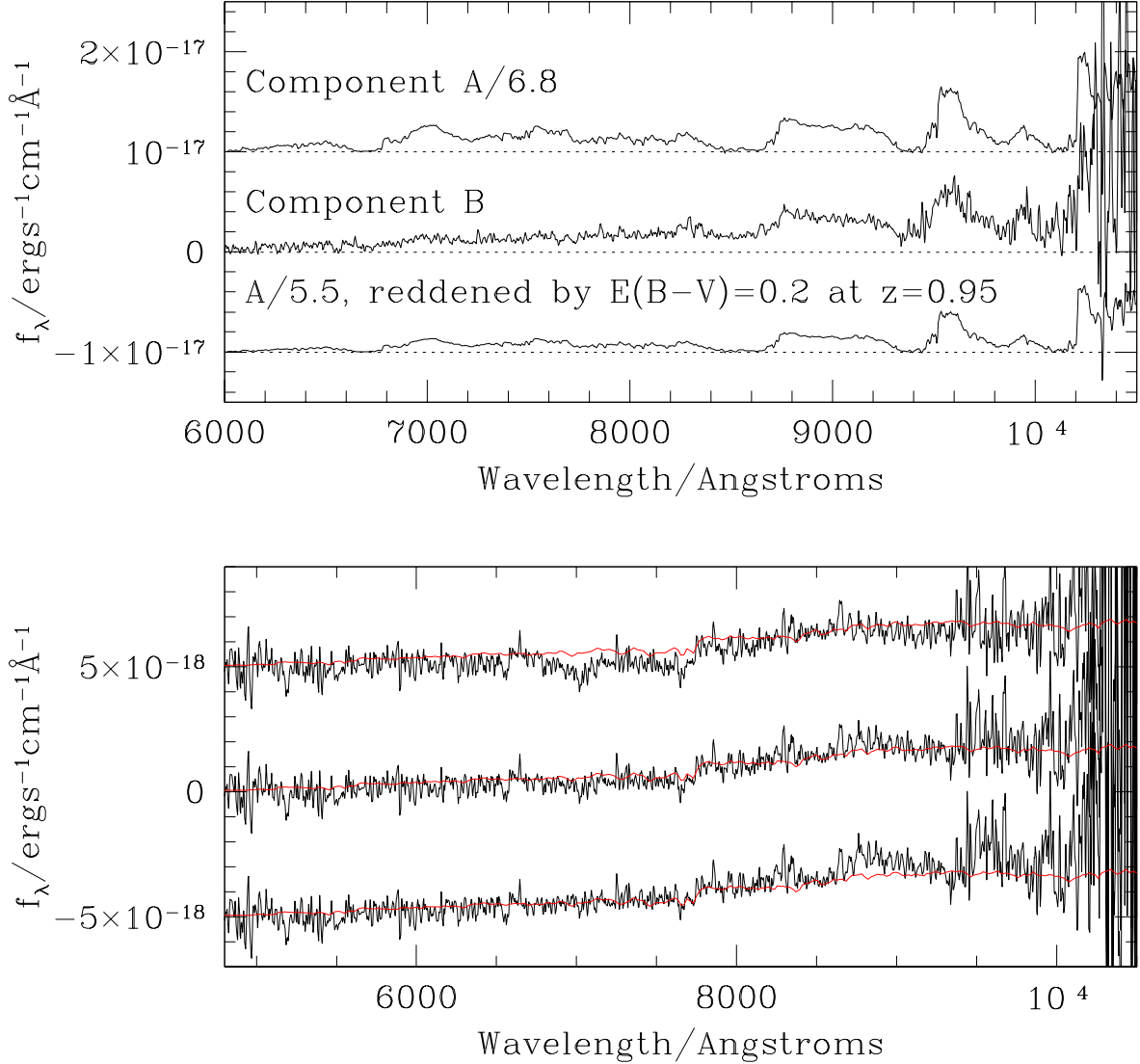}
\caption{Decomposition of the Keck ESI spectrum. Top panel: raw 
spectrum of component B (middle spectrum)
compared to the spectrum of A divided by the 
$K$-band flux ratio of 6.8 (upper spectrum) and the spectrum of A 
reddened by $E(B-V)=0.2$ at $z=0.95$, divided by 5.5 (lower spectrum). 
Lower panel: residuals obtained by subtracting from the spectrum of 
component B spectra 
of component A reddened in the rest-frame of the lens and normalized
to the $K$-band flux ratio. From top to bottom: A reddened by 
$E(B-V)=0.1$ and divided by 6.2; A reddened by $E(B-V)=0.2$ and divided
by 5.5, and A reddened by $E(B-V)=0.3$ and divided by 5. 
Superposed in red on all three spectra is a 6Gyr old, 1Gyr
burst galaxy model from the PEGASE library of Fioc \& Rocca-Volmerange 
(1997) redshifted 
to $z=0.95$, our best estimate of the lens redshift.}
\end{figure}


\section{Discussion}

\subsection{The dust in the lensing galaxy}

The low but non-zero value for the excess reddening produced by the 
lensing galaxy is at the upper end of the range observed for
local ellipticals (e.g.\ Goudfrooij \& de Jong 1995), and also 
those inferred for other quasars lensed by 
$0.3 \stackrel{<}{_{\sim}} z \stackrel{<}{_{\sim}}1$ 
elliptical galaxies (Falco et al.\ 1999).  Elliptical
galaxies at $z\sim 1$, however, are more likely to have low-level star
formation episodes occurring in their cores (Menanteau, Abraham \&
Ellis 2001), so it is possible that such galaxies would typically have
more dust in their inner few kpc than is usually seen either 
in local ellipticals or in $z\sim 0.5$ lensing galaxies.

\subsection{The intrinsic nature of the quasar}

LoBALs are in general redder than normal quasars, almost certainly
because of reddening in the quasar environment.  Estimating the
reddening of this quasar is difficult because of the strong absorption
features in the rest-frame UV.  If, however, J1004+1229 is like lower
redshift FeLoBALs there should be little line absorption longward of
Mg{\sc ii} 2798, and we can make a very rough estimate of the
reddening by comparing the observed optical to near-infrared spectral energy
distribution (SED) with that
expected from a normal FBQS quasar at this redshift.  We assume that
the effect of reddening in the lens is negligible compared to that in
the quasar environment: we expect dust only in the inner few kpc of
the elliptical, and, as discussed above, this has only a small effect
on the fainter image; the brighter image, with an impact parameter
$\sim 10$~kpc is unlikely to be significantly reddened by dust in the
lens.  In Figure~7 we plot a comparison of the SED of J1004+1229 with
that of FBQS composites (Brotherton et al.\ 2001) reddened with the
Small Magellanic Cloud (SMC) extinction law of Pei (1992), from which we
estimate an $A_V\sim 1$ mag.  This is similar to the $A_V \sim 1.5$
estimated for the FeLoBAL J1556+3517 by Najita, Dey \& Brotherton
(2001) (see also Fig.\ 1).

With a radio luminosity at 5GHz, $L_{\rm 5 GHz} \sim
10^{25} {\rm WHz^{-1}sr^{-1}}$ after correction for lensing, the
quasar is in the radio-intermediate class when this is defined through
radio luminosity.  In terms of the radio-to-UV continuum ratio $R^{*}$
(Sramek \& Weedman 1980), $R^{*}\approx 280$ before correction of the
UV flux for reddening and BAL absorption.  After correction by an
assumed $E(B-V)=0.35$ for reddening, and an estimated BAL absorption
of a further factor of two, this results in an $R^{*}\approx 10$,
again placing the quasar in the radio-intermediate class, as is typical
for BALs in FIRST (Becker et al.\ 2000; 2001).

At restframe optical wavelengths the extinction due to dust
approximately matches
the magnification due to gravitational lensing, and using this
assumption we can use the width of the H$\alpha$ line to estimate a
black hole mass of a few times $10^9M_{\odot}$, following the
prescription used by Lacy et al.\ (2001).  Even with this high black
hole mass, the high intrinsic luminosity of the quasar requires that
it is accreting near the Eddington limit.  The high black hole mass is
consistent with its high radio luminosity for a BAL quasar.

\begin{figure}
\plotone{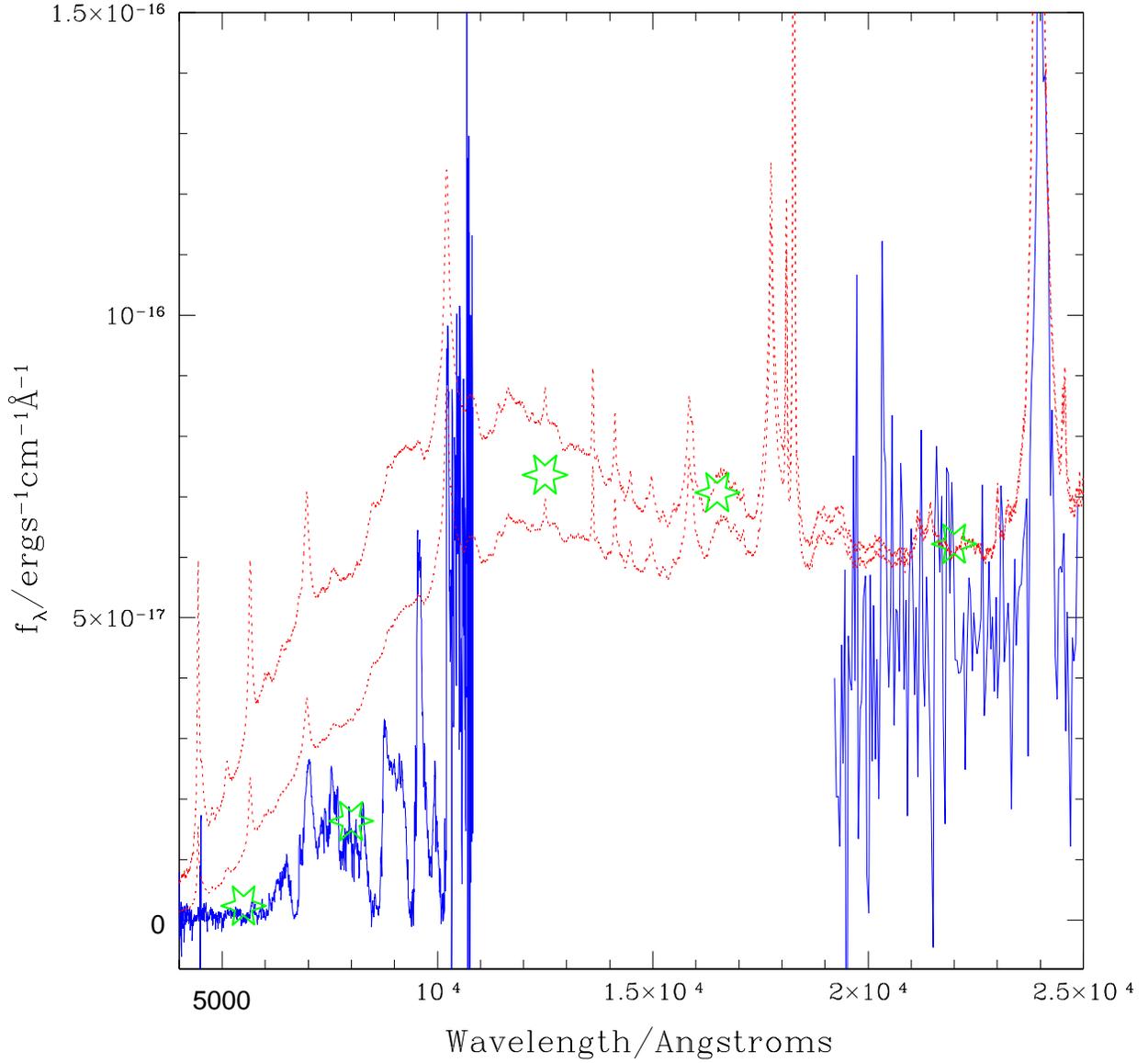}
\caption{The spectrum of J1004+1229A (solid blue line) compared to the FBQS
composite spectrum (Brotherton et al.\ 2001) reddened by $E(B-V)=0.3$
and $E(B-V)=0.4$ (upper and lower red dotted lines respectively). Points
from the 2-MASS $J$,$H$, $K$ and our NOT $V$ and $I$-band photometry
of the system are shown as green stars. The optical and infrared spectra
have been scaled to match the broad-band magnitudes in $I$ and $K$
respectively.}
\end{figure}

\section{Lensing of red quasars}

The steep number counts of high luminosity quasars produces a large 
magnification bias, i.e.\ a tendency for lensed quasars to be over-represented
in a flux-limited quasar survey (e.g.\ Turner, Ostriker \& Gott 1984). This 
typically results in such samples containing $\sim 1$\% of lensed 
quasars, compared to $\sim 0.1$\% expected in the absence of magnification 
bias. As our 2MASS quasar sample is very bright, the 
magnification bias is even higher than usual, and, using the
lensing model discussed below, we estimate that the 
expected lensing rate for $z>0.5$ quasars in our sample is
$\approx 3$\%. 
Even so, the discovery that 2/13 of our $z>0.5$ luminous red quasars 
are lensed is surprising. Although the numbers of quasars and lenses in our 
sample is small, 
the Poisson probability of finding two or more lensed quasars in 
our sample by
chance is only 0.06.  Why the lensed fraction is so high is not clear, but
it is probably linked to the observation of Chartas (2000) who showed
that the fraction of lensed quasars which are BALs is higher than that
in normal quasar samples.  Chartas argues that the fraction of
attenuated BAL quasars in samples of gravitationally-lensed quasars
can be explained as a result of the joint effects of lensing and
attenuation on the selection of these samples.  His analysis is
strongly dependent on the relative amounts of attenuation by dust and
magnification by lensing, in addition to the slope of the luminosity
function.  To see this, we follow Goodrich (1997) and Chartas (2000)
by considering a quasar population with a luminosity function
$\phi(L,z)$ studied in a narrow range of luminosity ($L_1,L_2$) at
redshift $z$. We draw samples of lensed quasars magnified by a mean
magnificaton factor $M$, and reddened quasars (including BALs),
attenuated by a mean attenuation 
factor $A$. We also now define the probability of
the quasar being lensed by $p_l$, and the probability of it being
significantly reddened by $p_r$.  The fraction of lensed quasars in
the sample is $f_l$:
\[ f_l = \frac{\Phi(L_1/M,L_2/M,z)}{\Phi(L_1,L_2,z)} p_l = b_l p_l\]
where $\Phi$ is the integral luminosity function and $b_l$ is the usual 
lensing magnfication bias factor (e.g.\ Turner, Ostriker \& Gott 1984). 
The fraction of red quasars in the sample is $f_r$:
\[ f_r = \frac{\Phi(L_1A,L_2A,z)}{\Phi(L_1,L_2,z)} p_r = b_r p_r\]
where we define $b_r$ as the ``attenuation bias factor'', by analogy
with $b_l$.  If the luminosity function is steep (i.e., equivalent to
a power-law of index $<-1$ over the luminosity ranges of interest),
then $b_l>1$ and $b_r<1$. The fraction of lensed quasars which are
reddened, $f_{lr}$ is then:
\[ f_{lr} = \frac{\Phi(L_1A/M,L_2A/M,z)}{\Phi(L_1/M,L_2/M,z)} p_r.\]
Hence if $A\approx M$ and the luminosity function is a single
power-law, then $f_{lr} = p_r / b_l \approx b_r p_r = f_r$, i.e.\ the
same as in the rest of the sample.  It is, however, suggestive in this
context that both the lensed objects are amongst the brightest in
$K$-band of all our FIRST-2MASS quasar candidates, where $b_l$ should
be highest. A steep cutoff in the luminosity function at high luminosities 
could have the effect of dramatically decreasing $b_r$ (as there are 
very few high luminosity quasars to attenuate) whilst keeping 
$b_l$ moderate. No evidence for such a cutoff has been found, however
(e.g.\ Wisotzki 2000).

We now make the discussion more quantitative, by considering
hypothetical quasar surveys selected in the $K$-band, with magnitude
limits of $K=15$, typical of the deeper portions of 2MASS, and $K=18$,
which would be typical of future large-area near-infared surveys with
4-m class telescopes, and $B$-band selected samples at $B=18$ and
$B=21$.  We adopt the luminosity function of Boyle et al.\ (2000)
\footnote{At $z\sim 2$ this luminosity function is consistent with
that of Wisotzki (2000), which, although it is only applicable to the
bright end, covers the range of magnitudes which includes our very
luminous quasars.}, and assume an optical spectrum whose flux density
$f_{\nu} \propto \nu^{-0.5}$.  We take the lensing probability
$p_l(z)$ from Kochanek (1993), and integrate over the magnification
distributions of King \& Browne (1996).  The distribution of
attenuations in the quasar population is uncertain.  The only sample
for which this has been well-studied is the extremely radio-bright 3C
sample, which, of course, may well not be typical of the quasar
population as a whole, and contains no BALs.  Simpson, Rawlings \&
Lacy (1999) find that, in a total sample of $\approx 32$ $z\sim 1$ 3C
radio sources, $28^{+25}_{-13}$\% of quasars are significantly
reddened with $A_V$ values ranging from $\approx 2$ to $\approx 15$.
Given the difficulty of defining a plausible attenuation distribution
we have decided to investigate how the fraction of reddened quasars in both
lensed and unlensed samples varies as a function of attenuation, using the 
SMC extinction law of Pei (1992).  In
Fig.\ 8, we plot the fraction of reddened quasars expected in
magnitude limited samples as a function of attenuation in the
rest-frame at $z=2$ in both lensed and unlensed samples, selected in
$K$-band and $B$-band.  The plots are normalized to unity
for zero attenuation.  Even for samples selected in the
near-infrared, red quasars with rest-frame $A_V \stackrel{>}{_{\sim}}1$
are strongly selected against. This is
particularly true in bright samples, where the slope of the luminosity
function is steepest.  Also, although the fraction of red quasars in
lensed samples is higher than that in unlensed samples, it remains a
factor of several lower than the true fraction.  These numbers depend
only on the magnitude limit of the quasar sample and the band to which
that limit is applied, and are independent of other selection criteria
which might further decrease the number of red quasars in a sample,
e.g.\ a blue color selection criterion.  Our model implies that the
high fraction of red quasars seen in lensed samples is not
likely to be entirely due to magnification bias effects, if the
assumption that reddened and unreddened quasars are drawn at random
from the same underlying population is correct.  In the
$B$-band, where quasar samples are typically selected, the selection
effects are even more pronounced, and in a $B\approx 18$
magnitude-limited sample (e.g.\ LBQS) very few $z>2$ quasars reddened by
$A_V>1$ in the rest-frame would ever be found.

\onecolumn

\begin{figure}[ht]
\plottwo{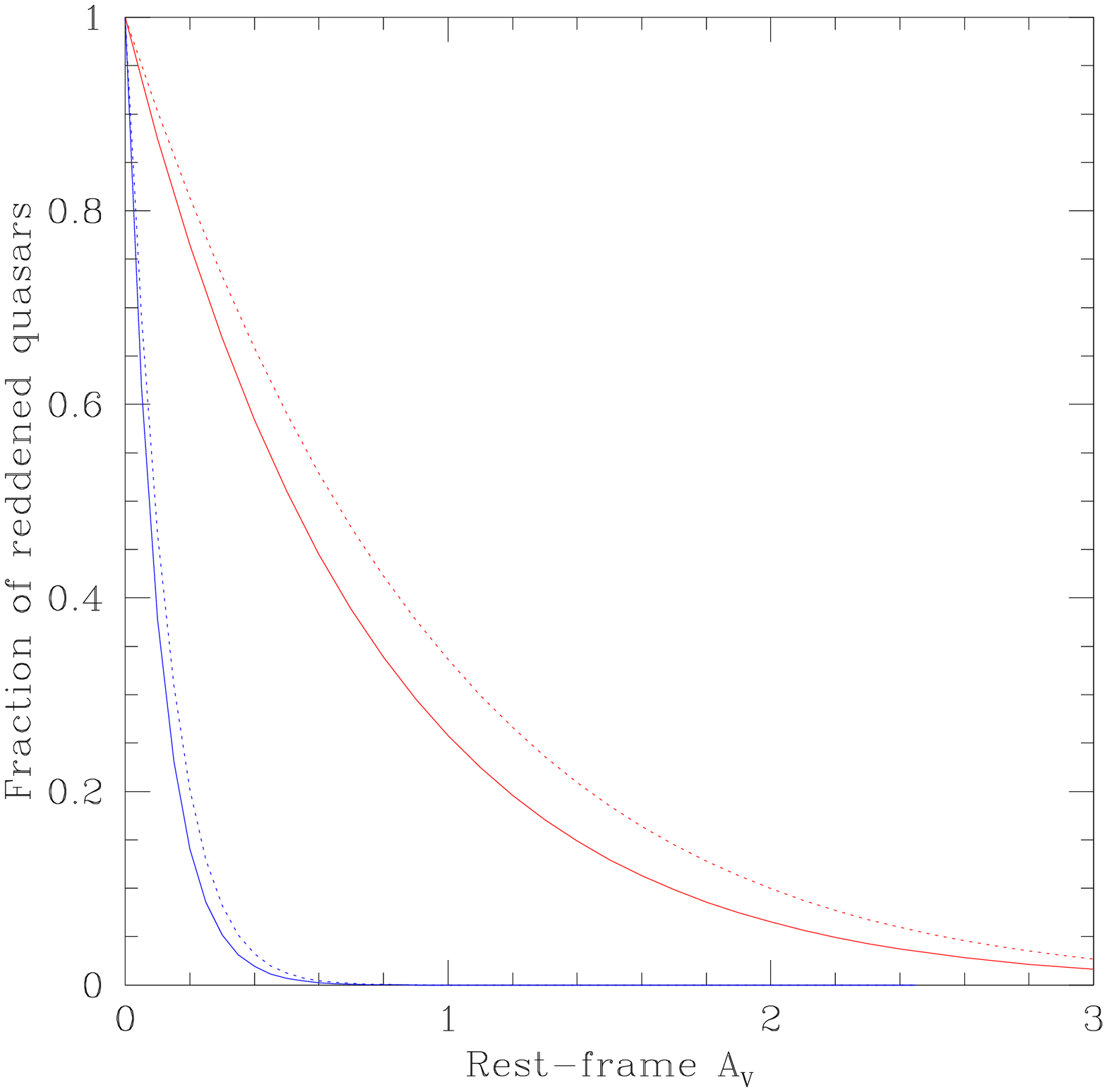}{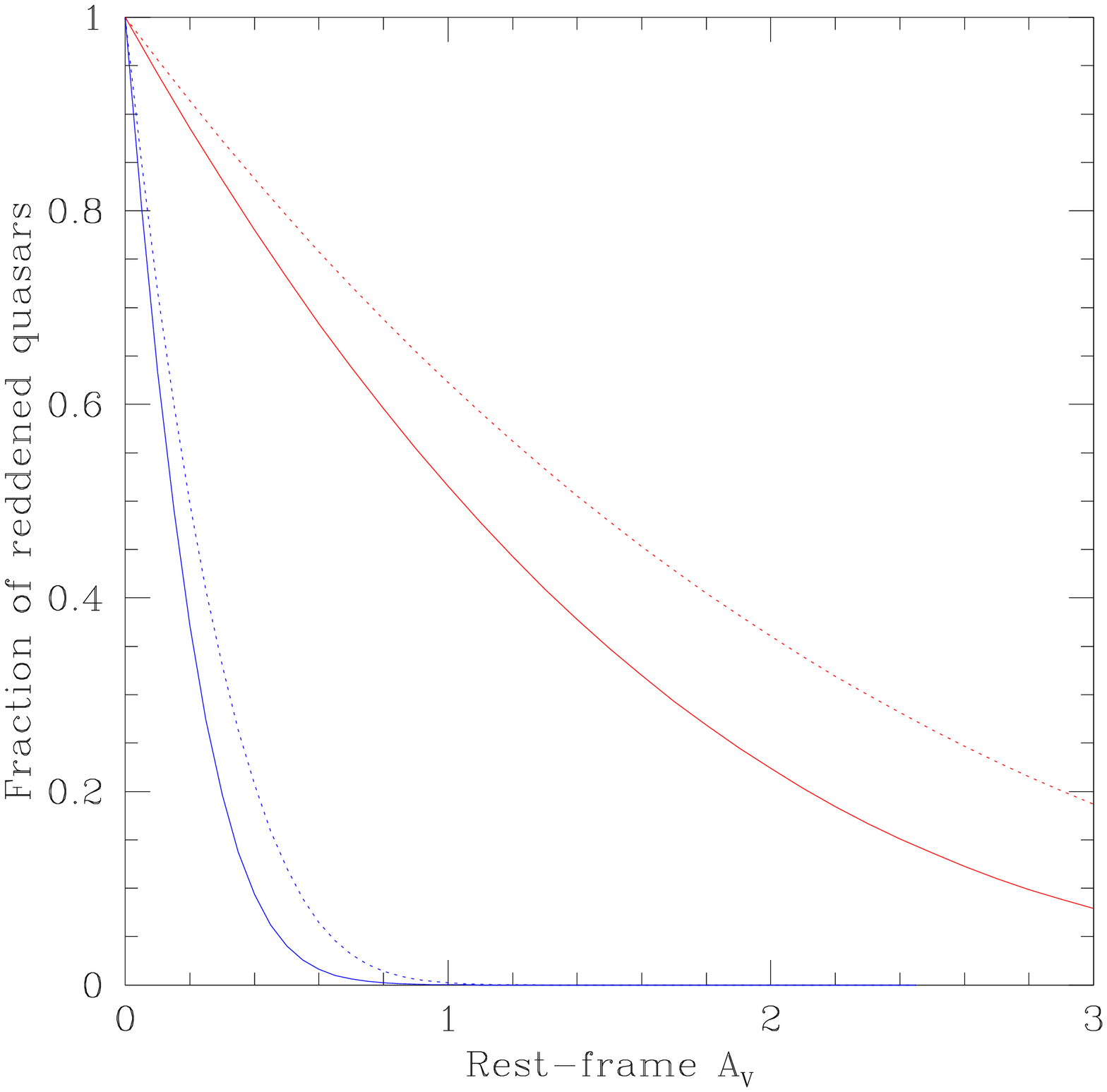}
\caption{The fraction of red quasars in 
magnitude-limited samples 
at $z=2$ as a function of rest-frame extinction. Left: 
the solid and dotted curves correspond to normal and lensed 
quasar samples in $K=15$ (red) and $B=18$ (blue) samples.
Right: the same for a sample with a $K=18$ (red) and a $B=21$ (blue)
magnitude limit. For comparison, J1004+1229 has $A_{V} \approx 1$ 
and J0134-0931 (lensed) and J0738+2927  (not lensed; 
Gregg et al.\ 2001) have $A_{V} \approx 2$.}
\end{figure}



An alternative explanation for the high fraction of red lensed quasars
is that red quasars are a distinct population, as might be the case if
redness is a phase in a quasar's life cycle rather than a purely
geometrical effect.  This is consistent with the interpretation of BAL
QSOs, especially the FeLoBALs, as being an early or heavily obscured
phase in the lives of quasars (Voit et al.\ 1993; Egami et al.\ 1996;
Becker et al.\ 1997, 2000; Gregg et al.\ 2000).  If the luminosity
function for red quasars is significantly steeper than for normal
quasars, then this could lead to a higher than normal magnification
bias in samples of red quasars, and explain our seemingly large
lensing rate, as well as the tendency for the fraction of BAL QSOs in
lensed samples to be higher.  A physical basis for this might be that
the luminosity function of quasars accreting at close to the Eddington
rate will resemble the black hole mass function, which has a steep
slope at high masses caused by the cutoff in galaxy masses produced by
the Schechter function (e.g.\ Salucci et al.\ 1999).  The only way for
a geometrical model to be consistent with this result is if the range
of angles over which the quasar is seen as significantly reddened
increases as quasar luminosity decreases.

\section{Conclusions}

The comparison of FIRST and 2MASS has allowed us to discover quasars
reddened by dust, both intrinsic and intervening.  J1004+1229 is
almost entirely reddened in the local environment of the quasar by a
combination of dust and strong low ionization absorption features; the
lens contributing only a small amount of reddening, and probably only
to the fainter lensed image.

Even after taking into account magnification by lensing, the quasar is
intrinsically very luminous and contains a supermassive black hole,
probably accreting at close to the Eddington rate. 
The apparently high
frequency of lensing among our red quasars could be explained in a
model in which red quasars have an intrinsically steeper luminosity function
than the normal quasar population, although if the geometry of the
reddening material is such that the range of angles in which reddening
is seen increases as quasar luminosity decreases, a geometrical
explanation could also work.  In either case, the extent to which 
high-redshift, dust-reddened quasars in both
lensed and unlensed are selected against in bright quasar
surveys is very great, and suggests that a large fraction of quasars
are missed from surveys with optical or near-infrared selection. In
particular it suggests that FeLoBALs, selected against both
because of dust reddening and extreme rest-frame UV absorption, may be
much more common than their rarity in magnitude-limited samples would
suggest.  When we complete our FIRST/2MASS survey, and analyse the
results in the context of our modelling we should obtain a good
estimate of the true fraction of red quasars in quasar samples of much
more moderate radio luminosity than those in the 3C survey.

The lensed nature of J1004+1229 gives us a unique opportunity to study
the host galaxy of a high-redshift BAL through deep imaging with
NICMOS on HST, or with ground-based adaptive optics.  If red quasars
and BALs are indeed a distinct and early phase in quasar evolution,
study of the host galaxy of J1004+1229 may help reveal the phenomena
responsible for creating such objects.

\acknowledgments

We thank Per Lilje for taking the $V$- and $I$-band images of 
J1004+1229 on the NOT, and Mike Brotherton for helpful discussions.
ML was a visiting Astronomer at the Infrared Telescope Facility,
which is operated by the University of Hawaii under contract
from the National Aeronautics and Space Administration. 
The NOT is operated on the island of La Palma jointly by
Denmark, Finland, Iceland, Norway, and Sweden, in the Spanish
Observatorio del Roque de los Muchachos of the Instituto de
Astrofisica de Canarias. ALFOSC is owned by the Instituto de
Astrofisica de Andalucia (IAA) and operated at the NOT under agreement
between IAA and the NBIfAFG of the Astronomical Observatory of
Copenhagen. The Two Micron All Sky Survey (2MASS) is a
joint project of the University of Massachusetts and the Infrared
Processing and Analysis Center/California Institute of Technology,
funded by the National Aeronautics and Space Administration (NASA) and the
National Science Foundation (NSF). This work was mostly performed under
the auspices of the U.S. Department of Energy,
National Nuclear Security Administration by the University of California,
Lawrence Livermore National Laboratory under contract No. W-7405-Eng-48,
with additional support from NSF grants AST-98-02791 (University of
California, Davis) and AST-98-02732 (Columbia University), with 
some work carried out at the Jet Propulsion Laboratory, California 
Institute of Technology, under contract with NASA.




\begin{thebibliography}{}
\bibitem[]{first} Becker, R.H., White, R.L., Helfand, D.J. 1995, ApJ,
450, 559
\bibitem[Becker et al.\ (1997)]{FBALS} Becker R.H., Gregg M.D., 
Hook I.M., McMahon R.G., White R.L., Helfand D.J., 1997, ApJ, 479, L93
\bibitem[Becker et al.\ (2000)]{RBALS} Becker R.H., White R.L., 
Gregg M.D., Brotherton M.S., Laurent-Muehleisen S.A., Arav N., 2000, 
ApJ, 538, 72
\bibitem[]{} Becker R.H., et al., 2001, ApJS, 135, 227
\bibitem[]{} Boyle B.J., Shanks T., Croom S.M., Smith R.J., Miller L., 
Loaring N., Heymans C., 2000, MNRAS, 317, 1014
\bibitem[Brotherton et al.\ 2001]{FBQScomp} Brotherton M.S., Tran H.D.,
Becker R.H., Gregg M.D., Laurent-Muehleisen S.A., White R.L., 2001,
ApJ, 546, 775  
\bibitem[]{} Browne I.W.A., et al., 1998, MNRAS, 293, 257
\bibitem[Canalizo \& Stockton 2001]{CS BAL} Canalizo G., Stockton A.N.,
2001, in, Crenshaw D.M., Kraemer S.B., George I.M., eds, 
Mass Outflows in Active Galactic Nuclei: New Perspectives". ASP 
Conf.\ Ser.\ in press (astro-ph/0107323) 
\bibitem[]{} Cardelli J.A., Clayton G.C., Mathis J.S., 1989, ApJ, 345, 245
\bibitem[Chartas 2000]{Ch00} Chartas G., 2000, ApJ, 531, 81
\bibitem[]{NVSS} Condon, J.J., Cotton W.D., Greisen E.W., Yin Q.F., 
Perley R.A., Taylor G.B., Broderick J.J., 1998, AJ, 115, 1693
\bibitem[Cowie et al.\ (1994)]{} Cowie L.L., et al., 1994, ApJ, 432, L83
\bibitem[Eales \& Rawlings 1993]{ER93} Eales S.A., Rawlings S., 1993, ApJ, 411,
67
\bibitem[]{} Egami, E., Iwamuro, F., Maihara, T., Oya, S., \& Cowie,
L. L. 1996, AJ, 112, 73
\bibitem[]{} Falco E., Impey C.D., Kochanek C.S., Leh\'{a}r J., McLeod B.A.,
Rix H.-W., Keeton C.R., Mu\~{n}oz J.A., Peng C.Y., 1999, ApJ, 523, 617
\bibitem[Fioc \& Rocca-Volmeragne 1997]{FRV97} Fioc M., Rocca-Volmerange
B., 1997, A\&A, 326, 950
\bibitem[]{} Gallagher, S.C., Brandt W.N., Chartas G., Garmire, G., 2001, 
ApJ, in press (astro-ph/0110579)
\bibitem[Goodrich 1997]{Go97} Goodrich R.W., 1997, ApJ, 474, 606
\bibitem[Goudfrooij \& de Jong 1995]{GdeJ95} Goudfrooij P., 
de Jong T., 1995, A\&A, 298, 784
\bibitem[]{} Gregg, M. D., Becker, R. H., Brotherton, M. S., Laurent-Muehleisen,
S. A., Lacy, M., \& White, R. L. 2000. 
ApJ, 544, 142
\bibitem[Gregg et al.\ (2001)]{0134} Gregg M.D., Lacy M., White R.L., 
Glikman E., Helfand D., Becker R.H., Brotherton M.S., 2001, ApJ, 564, 133
\bibitem[Hazard et al.\ (1987)]{HMW87} Hazard C., McMahon R.G., Webb J.K., 
Morton D.C., 1987, ApJ, 323, 263
\bibitem[]{} Kleinmann, S.G., Lysaght, M.G., Pughe, W.L., Schneider,
S.E., Skrutskie, M.F., Weinberg, M.D., Price, S.D., Matthews, K.,
Soifer, B.T., Huchra, J.P. 1994, ApJS, 217, 11
\bibitem[]{} King L.J., Browne I.W.A., 1996, MNRAS, 282, 67
\bibitem[]{} Kochanek C.S., 1993, MNRAS, 261, 453
\bibitem[Lacy et al.\ 2001]{L01} Lacy M., Laurent-Muehleisen S.A., 
Ridgway S.E., Becker R.H., White R.L., 2001, ApJ, 551, L17
\bibitem[Laor et al.\ 1997]{L97} Laor A., Fiore F., Elvis M., Wilkes B.J., 
McDowell J.C., 1997, ApJ, 477, 93
\bibitem[Mathur et al.\ 2001]{MMGES01} Mathur S., Matt G., Green P.J., 
Elvis M., Singh K.P., 2001, ApJ, 551, L13
\bibitem[Menanteau et al.\ (2001)]{MAE01} Menanteau F., Abraham R.G., 
Ellis R.S., 2001, MNRAS, 322, 1
\bibitem[]{} Menten K.M., Reid M.J., 1996, ApJ, 465, L99
\bibitem[]{} Merritt D., Ferrarese L., 2001, MNRAS, 320, L30
\bibitem[Najita et al.\ (2000)]{NDB00} Najita J., Dey A., Brotherton M.S.,
2000, AJ, 120, 2859 
\bibitem[]{} Pei Y.C., 1992, ApJ, 395, 130
\bibitem[]{} Salucci P., Szuszkiewicz E., Monaco P., Danese L., 1999, 
MNRAS, 307, 637; erratum MNRAS 311, 448
\bibitem[Simpson, Rawlings \& Lacy, 1999]{} Simpson C., Rawlings S., Lacy
M., MNRAS, 306, 828
\bibitem[Sprayberry \& Foltz 1992]{SF92} Sprayberry D., Foltz C.B., 1992, 
ApJ, 390, 39
\bibitem[]{} Sramek R.A., Weedman D.W., 1980, ApJ, 238, 435
\bibitem[TOG]{TOG} Turner E.L., Ostriker J.P., Gott J.R., 1989, ApJ, 284, 1
\bibitem[Voit et al.\ 1993]{VWK93} Voit G.M., Weymann R.J., Korista K.T., 
1993, ApJ, 413, 95
\bibitem[Weymann et al.\ (1991)]{WMFH91} Weymann R.J., Morris S.L., 
Foltz C.B., Hewett P.C., 1991, ApJ, 373, 23
\bibitem[]{} White, R.L., Becker, R.H., Helfand, D.J., Gregg,
M.D. 1997, ApJ, 475, 479
\bibitem[]{} Wiklind \& Coombes, 1996, Nat, 379, 139
\bibitem[Winn et al.\ 2001]{W0134} Winn J.N., Lovell J.E.J., Chen H.-W.,
Fletcher A.B., Hewitt J.N., Patnaik A.R., Schechter P.L., 2001, 
ApJ, 564, 143
\bibitem[]{} Wisotzki L., A\&A, 353, 853

\end{thebibliography}
\end{document}